\documentclass[%
 aip,
 amsmath,amssymb,
 reprint,%
]{revtex4-1}

\usepackage{graphicx}
\usepackage{dcolumn}
\usepackage{bm}
\usepackage[T1]{fontenc}  
\usepackage{textcomp}  
\usepackage{amsmath}
\newcommand{\angstrom}{\text{\normalfont\AA}}
\begin{document}


\title{Humidity effects on evolution of CsI thin Films: Fractal studies of rough surfaces}

\author{Nabeel Jammal}
\author{R. P.~Yadav}
\altaffiliation{Department of Physics, Government Post-Graduate. College, Saidabad, Allahabad-211004, India.}
\author{Triloki}
\altaffiliation[Present address: ]{Present address: INFN, Sezione di Trieste, Trieste, Italy.}
\author{R. Rai}
\author{A.K. Mittal}
\altaffiliation{Department of Physics, University of Allahabad, Allahabad, 211002, India.}
\author{B. K. Singh}%
\email{bksingh@bhu.ac.in.}
\affiliation{High Energy Physics Laboratory, Physics Department, Institute of Science, Banaras Hindu University, Varanasi 221005, India.}

\date{\today}

\begin{abstract}
The present work investigates the morphological, micro-structural, compositional and fractal analysis for CsI thin films in case of "as-deposited" and "1 hour humid air aged". The variation of grain sizes obtained from transmission electron microscopy (TEM) technique are found to be in the range of $\sim$ 313 nm to $\sim$ 1058 nm. The average grain size is found to increase after exposing to humidity. The experimental values of inter-planner spacing are found to be less than the standard value which signifies that a compressive stress is acting in the film. The elemental compositions of CsI film has been investigated by the means of energy dispersive X-ray spectroscopy (EDAX) technique. The atomic percentage of Cs:I (1:1) is found to be increased by a factor of two after exposing to humidity. The surface morphology of CsI thin films is analyzed by the atomic force microscopy (AFM) for both cases. The fractal analysis is performed on the AFM micrographs. The autocorrelation function and height-height correlation function is used to study the correlation properties of surface and roughness exponent. It is found that the fractal dimension is affected after exposing to humidity.
\end{abstract}

\maketitle

\section{\label{sec:level1}Introduction}

 The Cesium iodide (CsI) photocathodes as an efficient photo-emitter in the UV range have a great interest in high energy physics (HEP) and astrophysics research areas~\cite{639_(2011)_117-120,366_(1995)_60-70,371_(1996)_155,523_(2004)_345}. A flurry of devices in HEP experiments  such as; high acceptance di-electron spectrometer (HADES)~\cite{362_(1995)_183}, pioneering high energy nuclear interaction experiment (PHENIX)~\cite{646_(2011)_35} and a large ion collider experiment (ALICE)~\cite{ALICE}, etc, are employing CsI photocathodes for particle identification (PI). These high energy experiments require large area photocathode of few square meters across or more, operated in a high magnetic field. CsI photocathode can reach such dimensions and thus are very useful in these applications. Because of outstanding photoemissive and chemical properties, CsI films of transmission as well as of reflective mode are used extensively in solar blind photomultipliers as well as in Micro Channel Plate (MCP) based Vacuum Ultraviolet (VUV) and X-ray imaging detectors in astronomy~\cite{183_(1979)_169,219_(1984)_199}. CsI also has a variety of applications in other fields like medical imaging and particle physics experiment~\cite{44_(1994)_453,371_(1996)_137} due to its high quantum efficiency (QE) yield in the ultraviolet (UV) spectral range~\cite{14_(1975)_1667,297_(1990)_133,343_(1994)_99,343_(1994)_159}, a high secondary electron emission (SEE) yield~\cite{68_(1990)_2382,74_(1993)_7506}, a relatively high stability in air among other alkali halides material~\cite{New York_1980} and its capability to operate in a stable way with gaseous detectors~\cite{289_(1990)_322,(1989)_p._295}.
 
 Thin films surface morphologies control various essential physical and chemical properties of the films. Therefore it is extremely crucial to understand and control the evolution of the surface morphology during thin film growth~\cite{(1989)_p._295,117_(4)_2159-2166}. It is a complex phenomenon and very often occurs far from equilibrium~\cite{117_(4)_2159-2166,562_(2014)_126–131}. During thin film growth the atoms are deposited on surfaces which do not arrive uniformly at the surface. This random fluctuation, or noise, may create surface growth front roughness~\cite{94_ 042809_(2016)}. Surface roughness plays another important role in a number of physical phenomena including wave scattering~\cite{94_ 042809_(2016),Oxford_(1963),Publishing_(1991),Springer_(1994)}, friction~\cite{70_195409_(2004)}, adhesion~\cite{95_124301_(2005)}, electrical conductivity~\cite{61_11109_(1998),562_372376_(2014)}, capacitance~\cite{117_175308_(2015)}, heat transport~\cite{70_153404_(2004),79_1291_(2007)}, wetting~\cite{2016_120_5755−5763,122_185303_(2017)}, and sensors~\cite{71_155423_(2005),99_256101_(2007)}.
 
 The growth of thin films is a complicated stochastic process - its surface morphology depends on conditions of deposition, such as flux, angle of incidence and substrate temperature. The thin film surface plays a vital role in the interaction between the material and the environment. Therefore, its characterization is very important because each existing and prospective application of films strongly depend on the surface quality~\cite{117_(4)_2159-2166,562_(2014)_126–131}. The observed morphological properties like average roughness and interface width both depend on the resolution whereas they often scale in a resolution independent manner. Both average roughness and interface width are global measures of roughness but fail to describe the rough behavior on a nanometer scale, which is a topic of scientific interest~\cite{25_(2015)_083115,347_(2015)_706–712}. Growing surface is a complex phenomenon, which are far from equilibrium and naturally evolve into self-affine structures~\cite{117_(4)_2159-2166,562_(2014)_126–131}. The scaling concept is reminiscent of a fractal, which is used to describe self-affine surfaces. Fractal objects are characterized by only one scaling exponent and used to describe the complexity of thin films surfaces~\cite{117_(4)_2159-2166,562_(2014)_126–131,122_185303_(2017)}.
 
 In this paper, we investigated the morphological, micro-structural, compositional and fractal analysis for CsI thin films in case of "as-deposited" and "1 hour humid air aged". The surface heights of "as-deposited" and "1 hour humid air aged" thin films are measured using atomic force microscopy (AFM). The grain size was calculated by TEM technique. It is found that interface width, lateral correlation length, and roughness exponent increase and fractal dimension is influenced after exposing to humidity. 

\section{\label{sec:level1}Experimental Details}
The deposition of CsI thin films are carried out by the thermal evaporation technique under high vacuum ($7.3\times10^{-7}$ Torr). The complete details about the deposition processes are reported by Nabeel Jammal et al.~\cite{546_(2018)_21-27}. The prepared films have thicknesses of 500 nm, 300 nm, 50 nm and 30 nm with deposition rate $\sim1~nm/s$. After deposition, the "as-deposited" CsI thin films are extracted to a vacuum desiccator under constant flow of nitrogen (N$_2$) gas to avoid any interactions with the humidity. The vacuum desiccator, containing fresh silica gel, is then pumped for five minutes by a small diaphragm pump before its transported to the characterization setup.

A FEI-Technai 20 $G^2$, Transmission Electron Microscope (TEM) operated at an accelerating voltage of 200 kV is used in the current study. The TEM chamber is always evacuated to low pressure ($10^{-6}$~Torr) before performing any imaging to avoid interactions of high energy electrons with gas atoms which may lead to some sparks.  Moreover, TEM technique can focus the electron beam on a small area and it is possible to study a selected area of the sample. The Selected Area Electron Diffraction (SAED) pattern gives a direct indication of sample's crystallinity. The basic disadvantage of TEM technique is that it requires a special sample preparation as it puts a limitation on the sample's thickness. The sample must be thin enough so that it does not absorb the electrons and allows them to pass. Therefore, we have investigated the CsI thin film only with thickness of 100 nm deposited on copper grid substrate. Energy Dispersive X-ray Spectroscopy (EDAX) is a powerful technique that is ideal for revealing what elements are present in a particular sample. The instrument used for EDAX technique is attached with TEM FEI-Technai 20 $G^2$ technique. Atomic force microscope (AFM) scanning is done by solver-NEXT NT-MDT coupled with the PX Ultra controller available at Chemistry Department, BHU Varanasi. It provides a high resolution two dimensional (2D) and three dimensional (3D) surface images. In this study, CsI films have been deposited on S.S substrates. For each characterization, two samples are prepared; one of them is scanned just after deposition and the other exposed to the humidity for one hour before scanning. The topological scale (of 5 $\mu$m $\times$ 5 $\mu$m) has been used to investigate the surface morphology in AFM technique. The humidity is kept about $RH = 60\pm5\%$.
 
\section{\label{sec:level1}Results and discussion}
\subsection{\label{sec:level2}Morphological and compositional analysis}
The surface morphology obtained by TEM technique of "as-deposited" and "1 hour humid air aged" 100 nm thick CsI film is shown in Fig. 1. It is observed that the CsI thin film has homogeneous and continuous grain like morphology. The size of each grain is measured by using the scale length shown in Fig. 1. The average grain size of a specific image is calculated from the grain size distributions. The obtained grain size is found to be in the range of $\sim$ 313 nm to $\sim$ 1058 nm. In the case of "as-deposited", the average grain size is found to be $\sim$ 679 nm. In the case of "1 hour humid air aged", the average grain size has been increased due to the coalescence processes and is found to be $\sim$ 724 nm.

\begin{figure}[h]
	\begin{center}
		\includegraphics[width=\columnwidth,height=5cm]{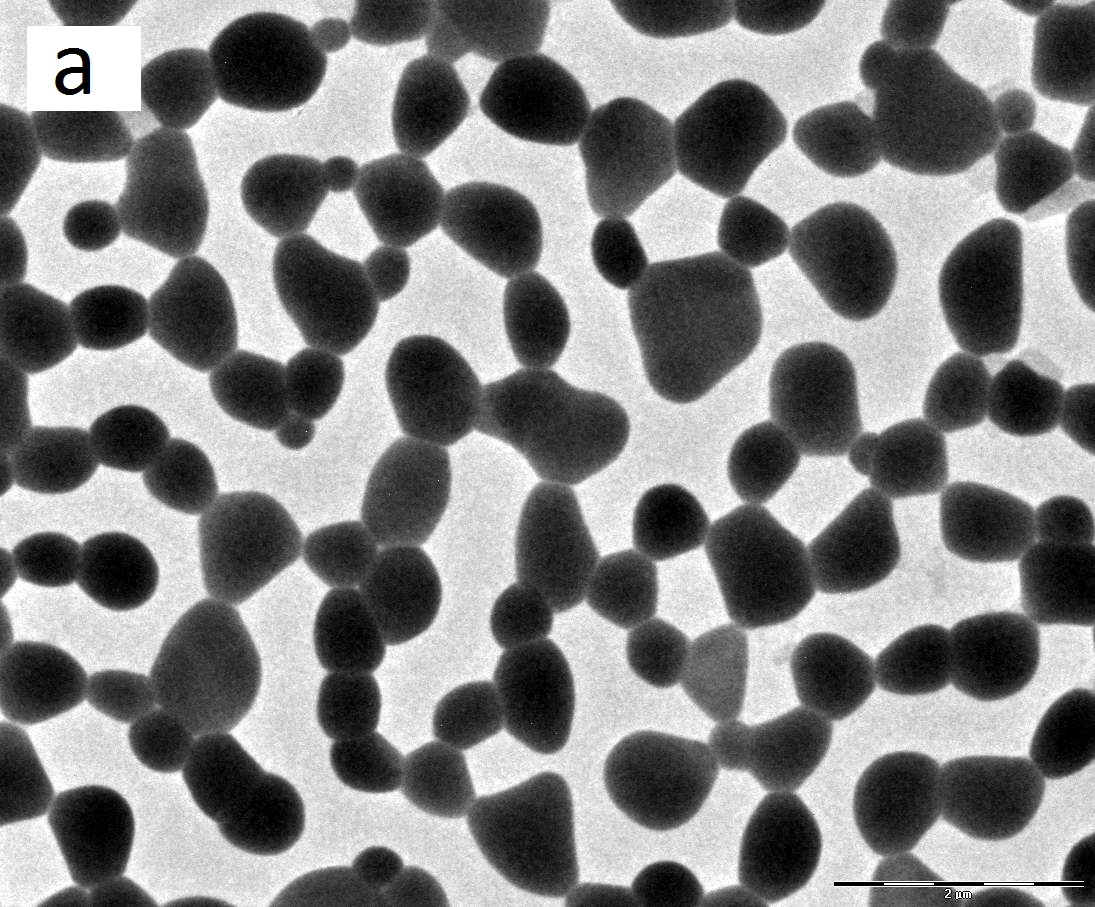}
		
		\vspace{0.5cm}
		\includegraphics[width=\columnwidth,height=5cm]{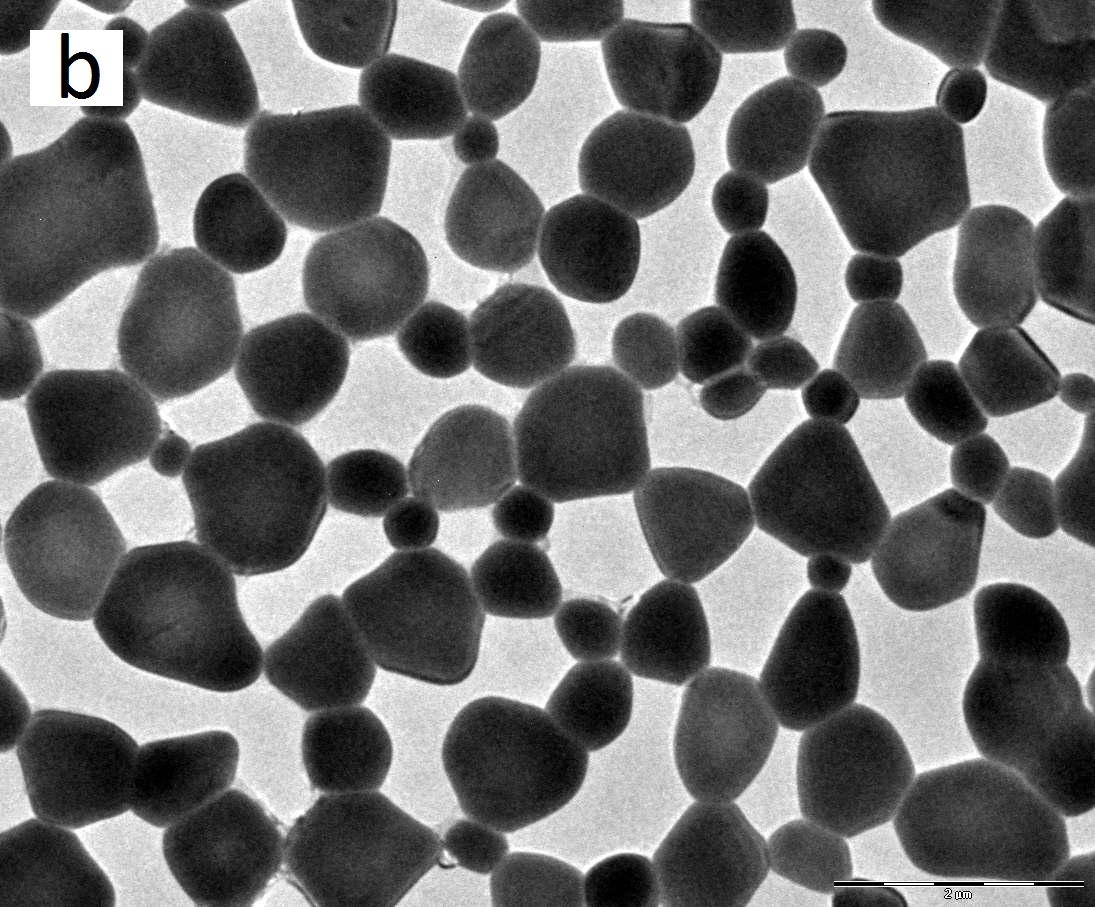}
		
	\end{center}
	\caption{The surface morphology of 100 nm thick CsI films in case of (a): "as-deposited" and (b): "1 hour humid air aged" obtained by TEM technique.}
\end{figure}

Generally, TEM technique is used to determine the grain size and to study the crystallinity of the films. The structural characterization of CsI thin film has been analyzed by studying the Selected Area Electron Diffraction (SAED) patterns in case of "as-deposited" and "1 hour humid air aged" as shown in Fig. 2. It can be noticed that the SAED patterns contain a huge number of spots; each spot arises from Bragg's reflection from an individual crystallite. The electron diffraction pattern shown in the figure indicates that CsI thin films are crystalline in nature and the separated lattice of sharp spots indicates that the films have single-crystal like domains. In case of "as-deposited" 100 nm thick CsI film, it is found that the radius of the first ring is about 3.18 (1/nm) and by matching with the previously available values from JCPDS pdf no.: 060311, it is found to be for the (hkl) lattice plane (110) and the lattice constant is found to be 4.444 $\angstrom$. Subsequent measurements are done for other radii and the (hkl) values along with the lattice constants and the stress obtained are tabulated in Table 1. One can observe that the obtained lattice planes from CsI thin films are attributed to the body centered cubic (bcc) structure. The values of lattice constant are in the range of 4.152 $\angstrom$ - 4.613 $\angstrom$ which is very close to the standard value of 4.567 $\angstrom$ reported in JCPDS pdf no.: 060311. Further, one can observe that the inter-planner spacing decreases with respect to the standard value which implies a compressive stress acting in the films.

\begin{figure}[!h]
	\centering
	\includegraphics[width=\columnwidth,height=5cm]{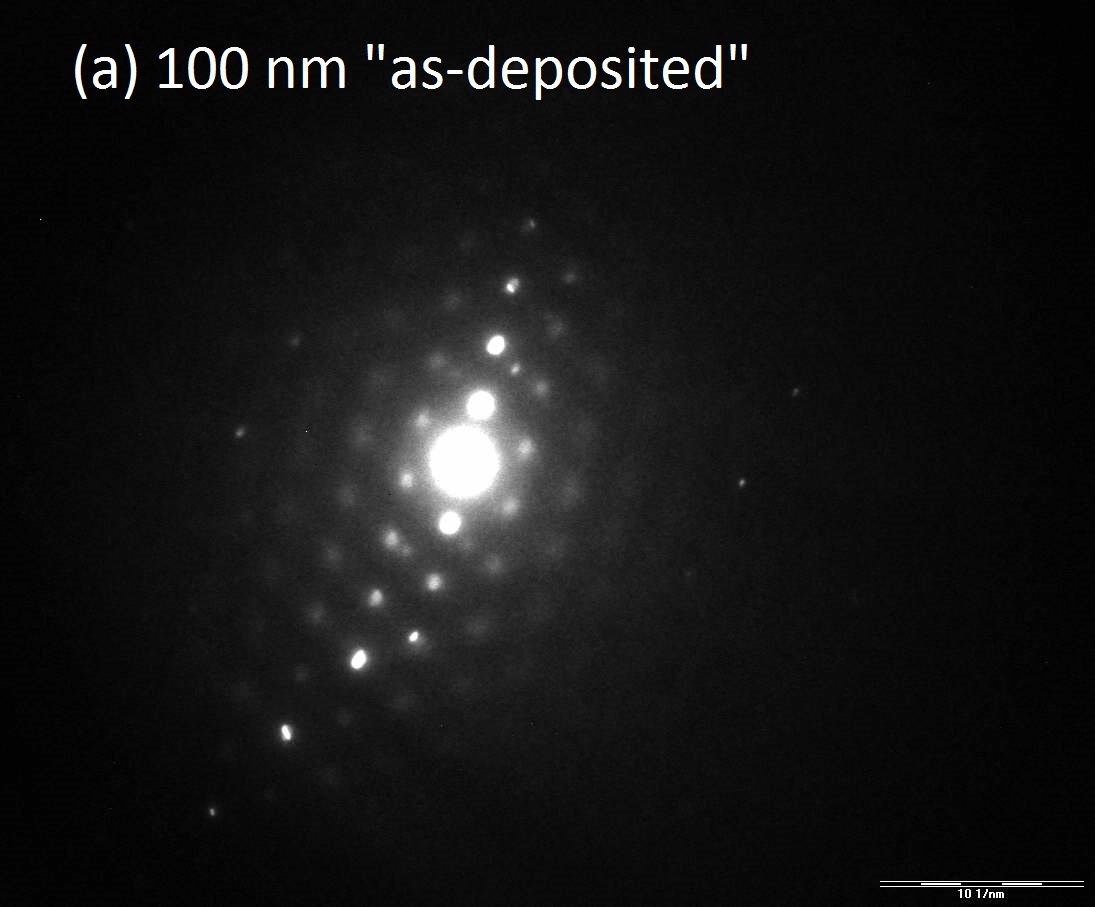}
	
	\vspace{0.5cm}
	\includegraphics[width=\columnwidth,height=5cm]{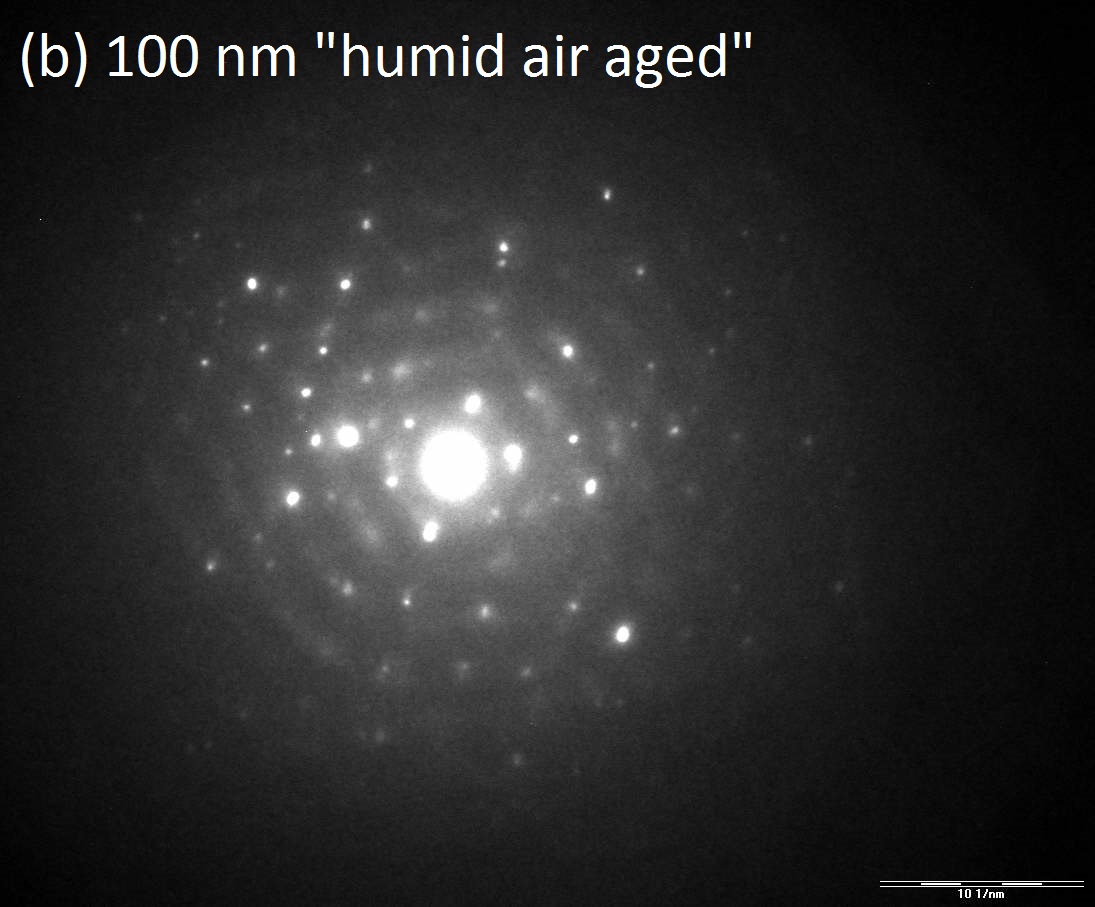}
	
	\caption{The SAED patterns of 100 nm thick CsI films in case of  (a): "as-deposited" and (b): "1 hour humid air aged".}
\end{figure}

\begin{table}[h]
	\begin{center}
		\caption{Experimental and calculated values of structural parameters in case of (a): "as-deposited" and (b): "1 hour humid air aged" 100 nm thick CsI film analyzed by TEM technique.}	 
		~\\   
		\begin{tabular}{c c c c c c c c} 
			\hline
			& Ring Radius & d$_{(exp)}$ & d$_{(stnd)}$ & (hkl) & a & $\Delta$ d & $\Delta$ d/d \\
			[0.5ex]
			& (1/nm) & ($\angstrom$) & ($\angstrom$) & ~ & ($\angstrom$) & ($\angstrom$) & ~
			\\[0.5ex]
			\hline
			& 3.18 & 3.144 & 3.230 & (110) & 4.444 & 0.086 & 0.02735\\
			(a) & 5.36 & 1.863 & 1.865 & (211) & 4.613 & 0.002 & 0.00011\\
			& 9.63 & 1.038 & 1.142 & (400) & 4.152 & 0.040 & 0.03200\\
			&11.30 & 0.884 & 0.896 & (510) & 4.507 & 0.012 & 0.01357\\		
			\hline
			& 3.17 & 3.155 & 3.230 & (110) & 4.460 & 0.0750 & 0.0240\\
			(b) & 5.38 & 1.859 & 1.865 & (211) & 4.554 & 0.0006 & 0.0003\\
			& 7.86 & 1.272 & 1.319 & (222) & 4.410 & 0.0047 & 0.0370\\
			&11.29 & 0.886 & 0.896 & (510) & 4.517 & 0.0100 & 0.0113\\		
			\hline
			
		\end{tabular}
	\end{center}
\end{table} 

Fig. 3 shows the elemental compositional analysis of 100 nm thick CsI film in case of "as-deposited" and "1 hour humid air aged" recorded by the EDAX technique. The spectrum peaks revealed the presence of cesium (Cs) and iodide (I) elements in the grown CsI films and some other peaks such as carbon (C), copper (Cu) and oxygen (O) are also observed. The copper and carbon peaks are seen to be originated from the copper grid that is used for mounting the sample in TEM technique. The elemental compositions corresponding to the "as-deposited" and "1 hour humid air aged" are also compared in Table 2. It can be noticed that the atomic percentage ratio of Cs to I is 24.70\% : 24.30\% in case of "as-deposited" and 50.40\% : 49.60\% in case of "1 hour humid air aged". Thus we may say that the ratio of Cs:I is $\sim$ 1:1 for both cases. Comparable result has been obtained recently by Triloki et al.~\cite{785_(2015)_70-76} for 500 nm thick CsI film. Further, the atomic percentage of both elements is found to be increased by a factor of two after exposing to humidity. This can be attributed directly to the increase in grain size with the humidity.

\begin{figure}[!h]
	\centering
	\includegraphics[width=\columnwidth,height=5cm]{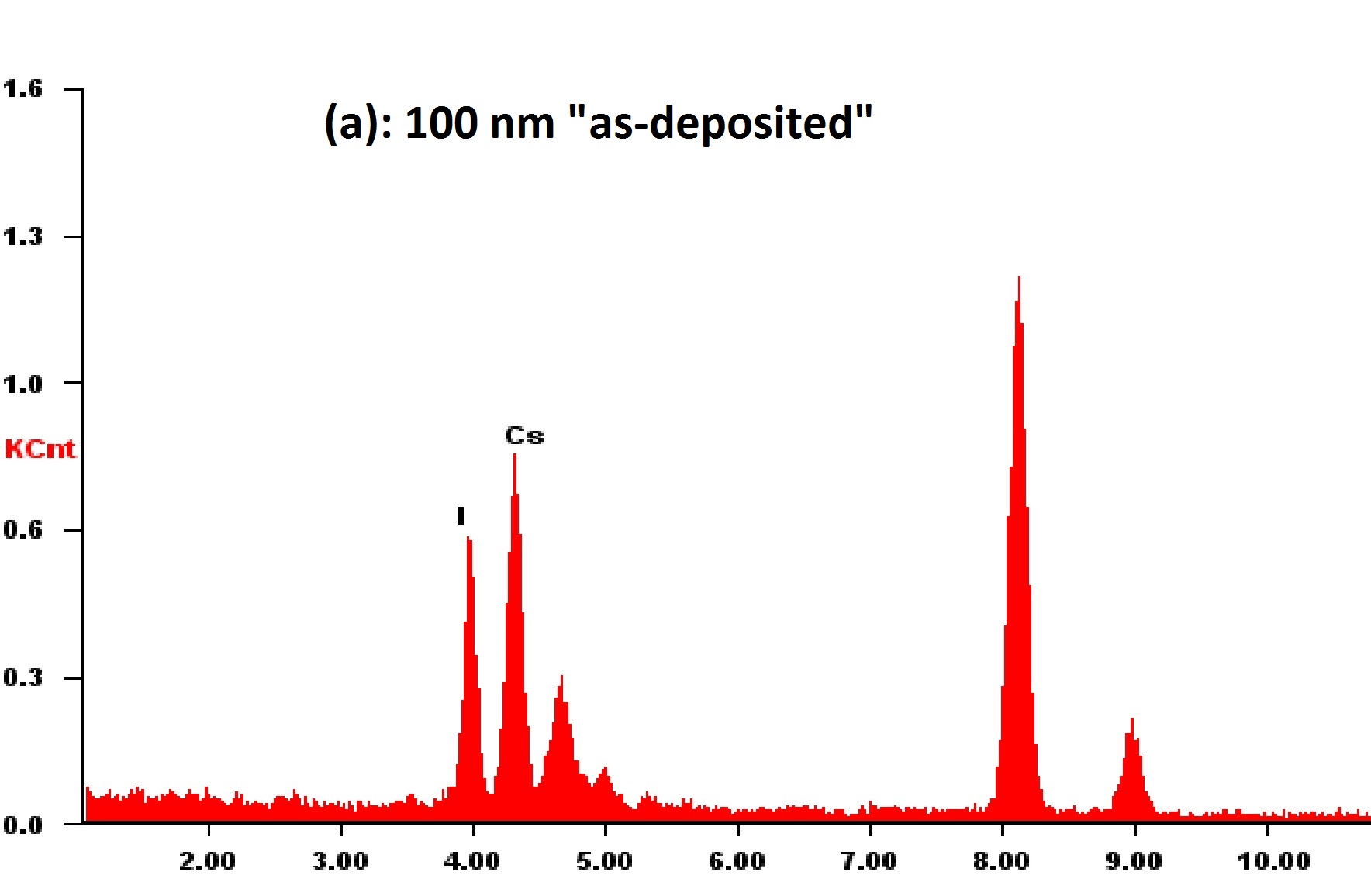}
	
	\vspace{0.5cm}
	\includegraphics[width=\columnwidth,height=5cm]{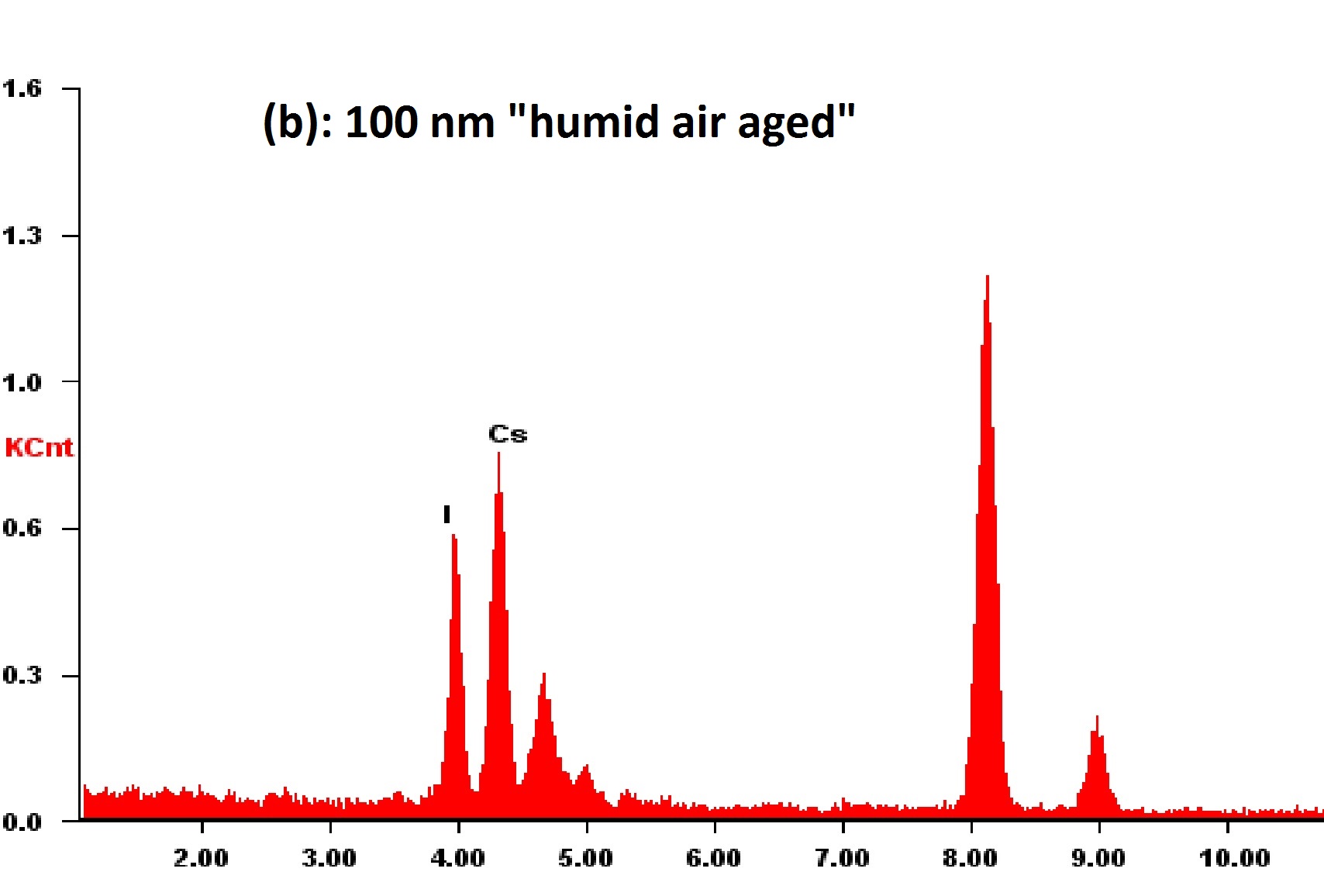}
	\caption{The EDAX patterns of 100 nm thick CsI films in case of "as-deposited" and "1 hour humid air aged".}
\end{figure}

\begin{table}[h]
	\begin{center}
		\caption{The EDAX data of 100 nm thick CsI films in case of (a): "as-deposited" and (b): "1 hour humid air aged".}	 
		~\\   
		\begin{tabular}{c c c c c} 
			\hline
			Element & Weight\% & Atomic\% & Weight\% & Atomic\% \\
			[0.5ex]
			~ & (a) & (a) & (b) & (b)
			\\[0.5ex]
			\hline
			Cs (L) & 46.40 & 24.70 & 51.60 & 50.40 \\
			I (K) & 43.60 & 24.30 & 48.40 & 49.60 \\		
			\hline
			
		\end{tabular}
	\end{center}
\end{table} 

\subsection{\label{sec:level2}Surface Roughness and Fractal analysis}
The surface morphology obtained from AFM technique of "as-deposited" and "1 hour humid air aged" CsI thin films are shown in Figs. 4 and 5. It can be observed that the grain density is increased with the film thickness which clearly inferred that the granules of various scales, irregular shapes, sizes and separations are present in the thin films. The smaller grains are seen at lower thickness which aggregate to form bigger grains at higher thickness. These surface structure parameters promote the surface roughness. A close and careful visualization of the AFM images suggest that "as-deposited" thin films [Fig.4 (a, c, e, g)] have more number of the kinks/spikes as compare to "1 hour humid air aged" thin films as shown in Fig. 4(b, d, f, h). 

\begin{figure}[!h]
	\begin{center}
		\includegraphics[scale=0.18]{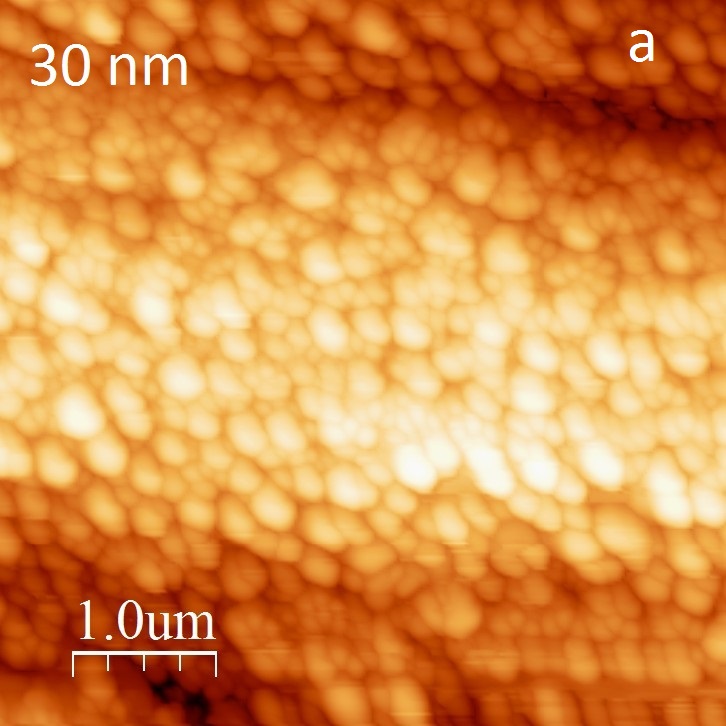}
		\vspace{0.1cm}
		\includegraphics[scale=0.18]{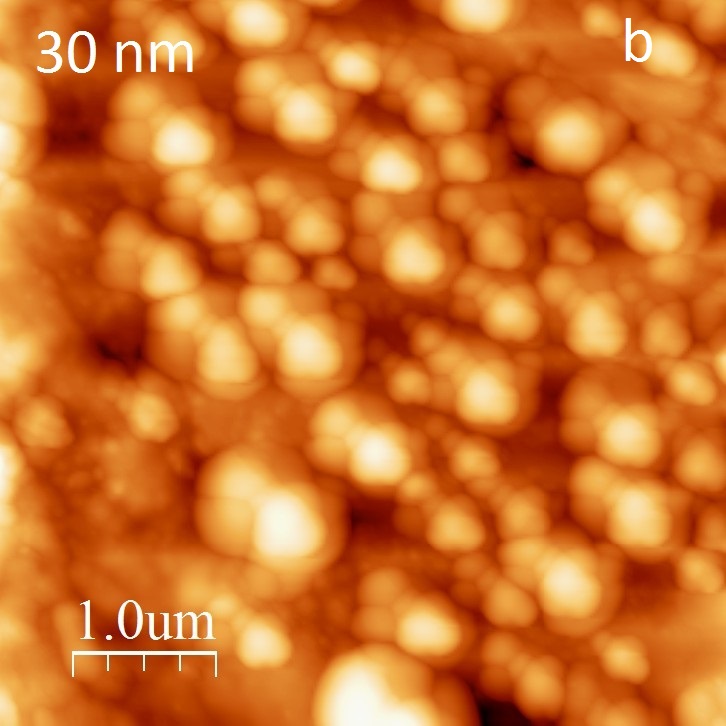}
		\vspace{0.1cm}
		\includegraphics[scale=0.18]{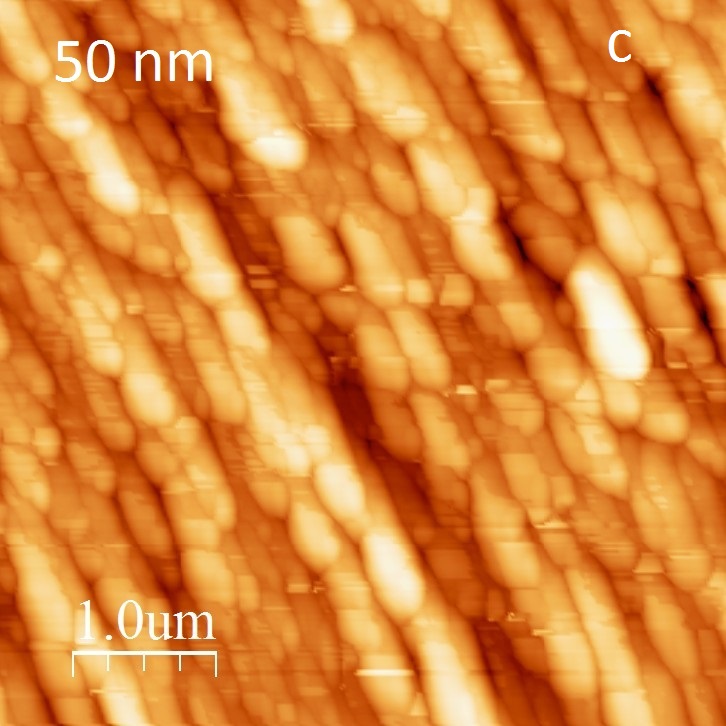}
		\vspace{0.1cm}
		\includegraphics[scale=0.18]{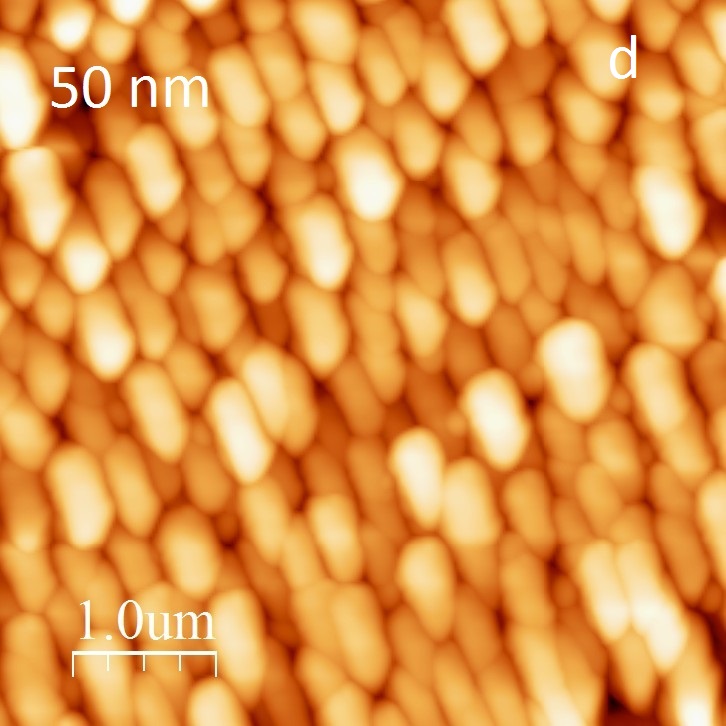}
		\hspace{0.1cm}		
		\includegraphics[scale=0.18]{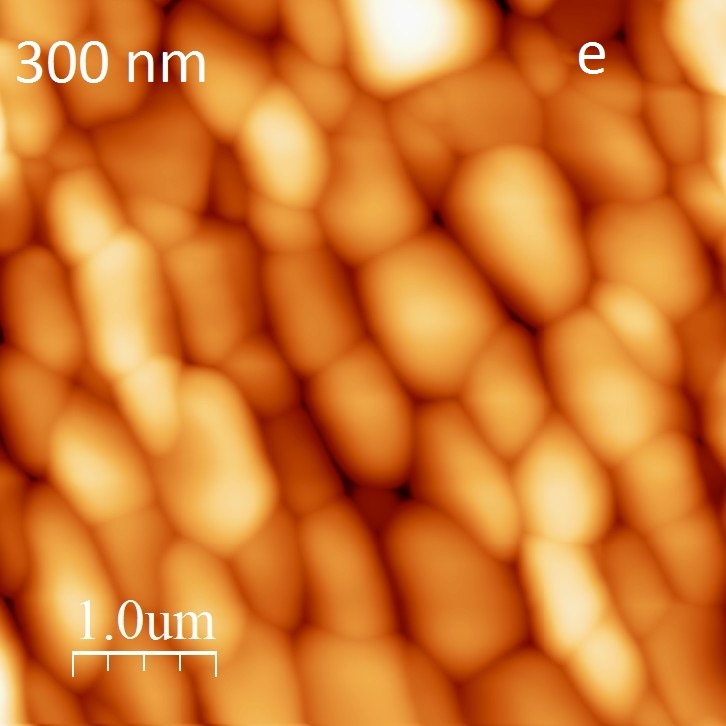}
		\vspace{0.1cm}
		\includegraphics[scale=0.18]{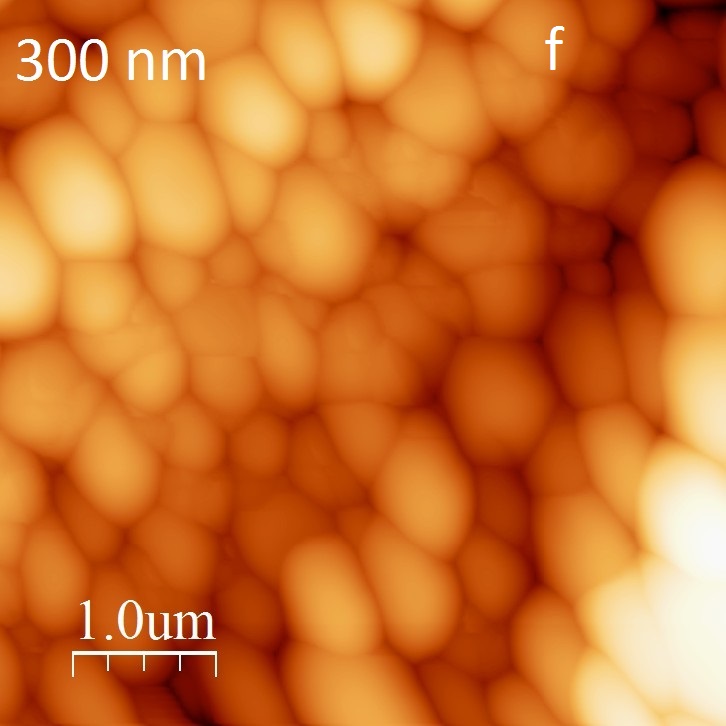}
		\vspace{0.1cm}
		\includegraphics[scale=0.18]{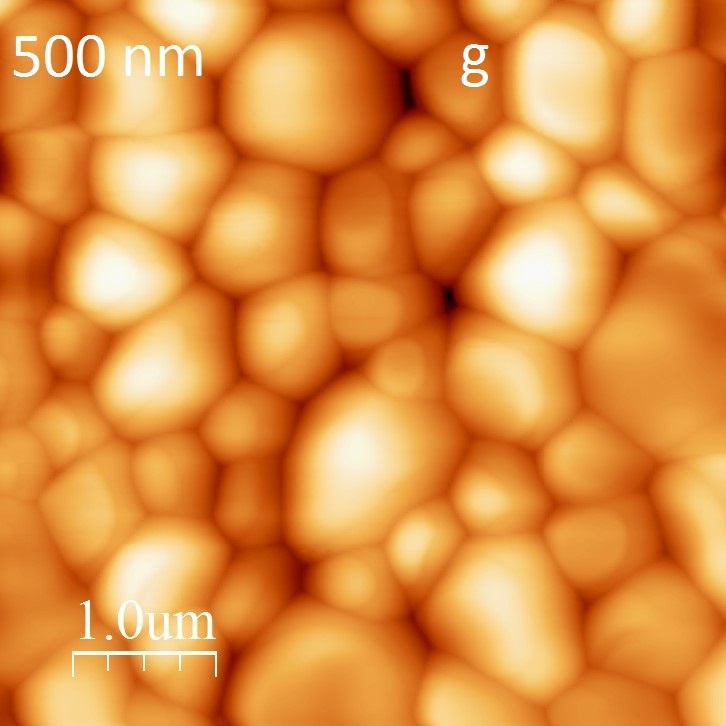}
		\vspace{0.1cm}
		\includegraphics[scale=0.18]{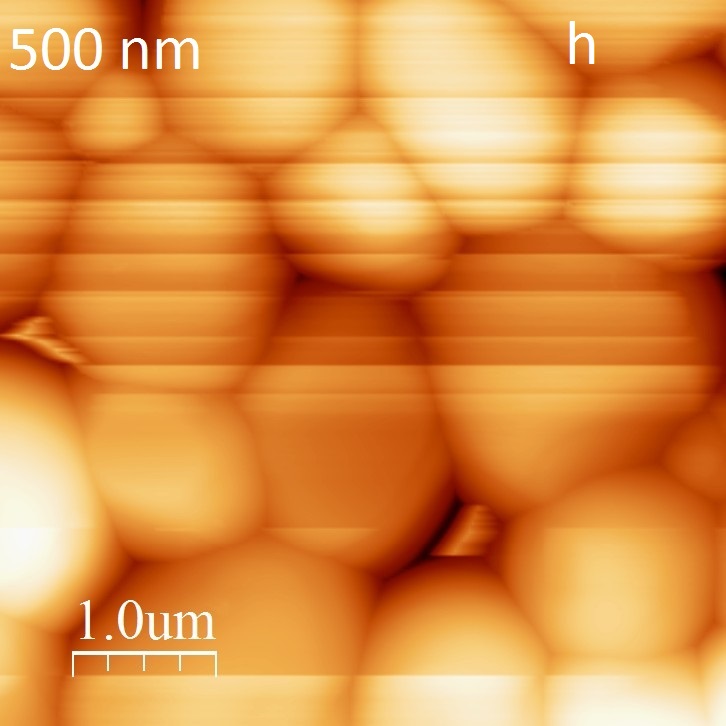}
	\end{center}
	\caption{The 2-D surface morphology of CsI thin films obtained by AFM technique in case of "as-deposited" (a, c, e, g) and "1 hour humid air aged" (b, d, f, h).}
\end{figure}

\begin{figure}[!h]
	\begin{center}
		\includegraphics[scale=0.17]{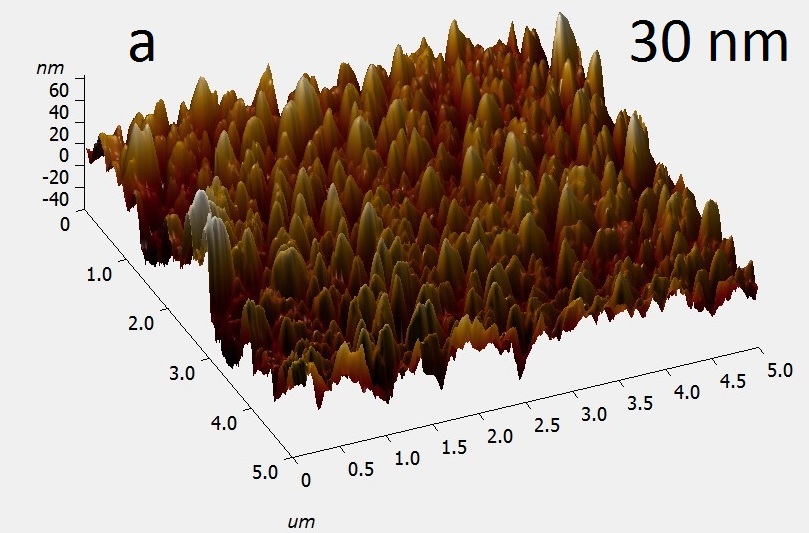}
		\vspace{0.1cm}
		\includegraphics[scale=0.17]{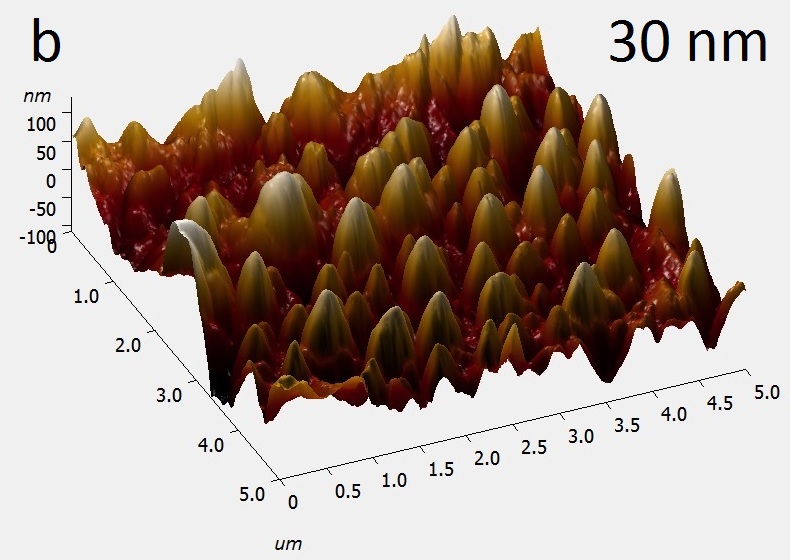}
		\vspace{0.1cm}
		\includegraphics[scale=0.17]{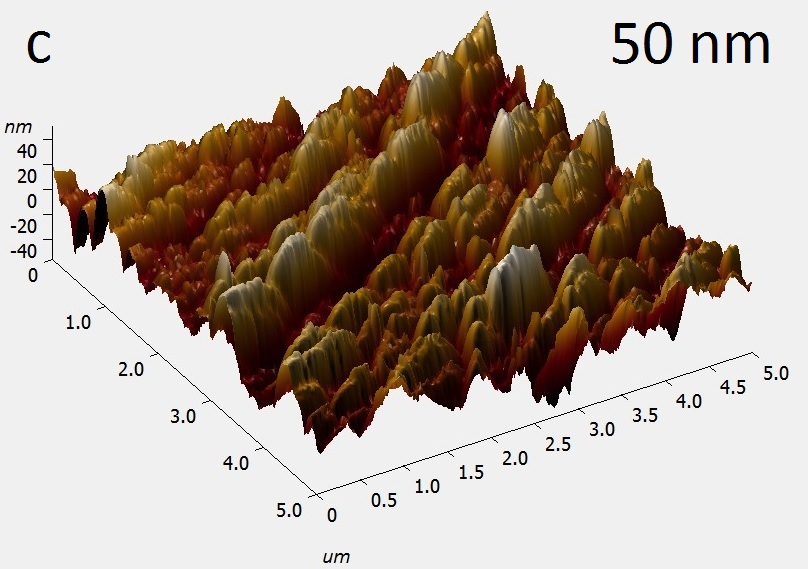}
		\vspace{0.1cm}
		\includegraphics[scale=0.17]{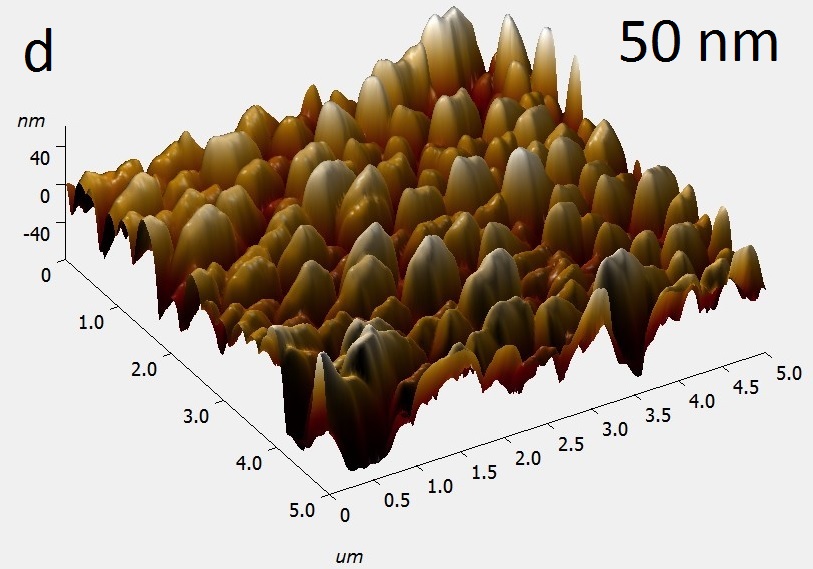}
		\hspace{0.1cm}		
		\includegraphics[scale=0.165]{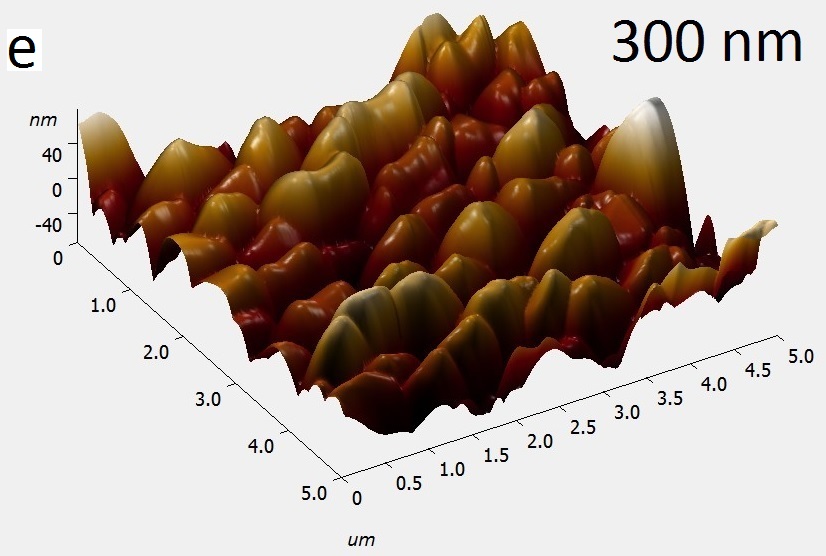}
		\vspace{0.1cm}
		\includegraphics[scale=0.165]{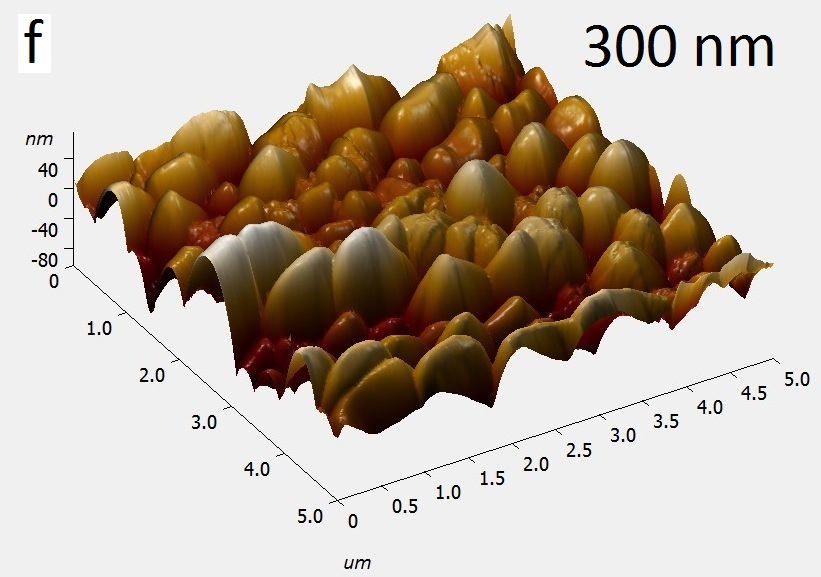}
		\vspace{0.1cm}
		\includegraphics[scale=0.165]{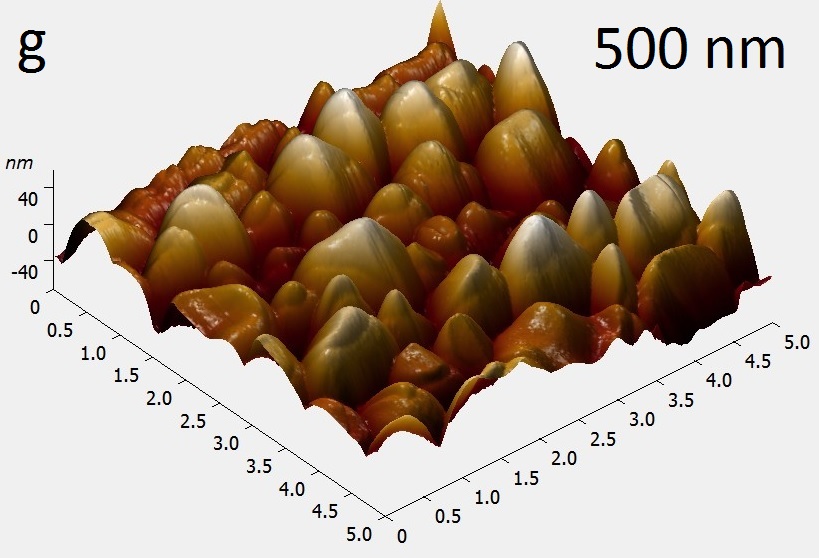}
		\vspace{0.1cm}
		\includegraphics[scale=0.165]{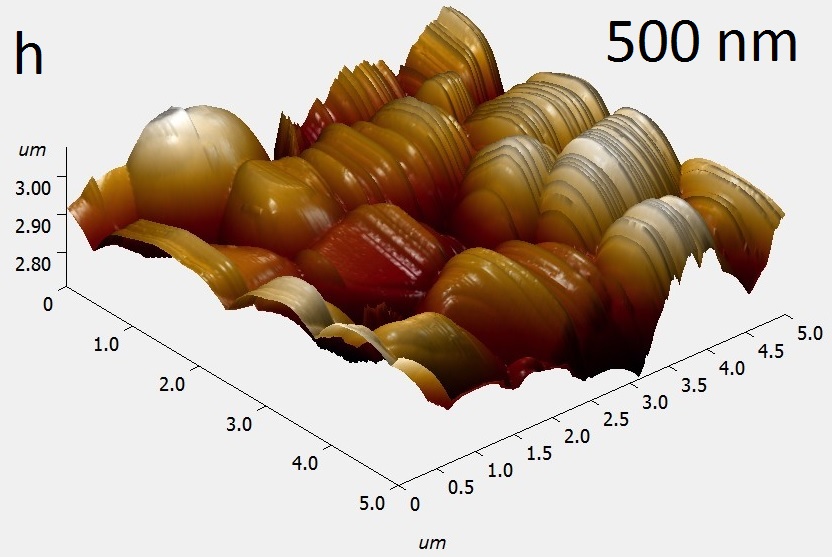}
	\end{center}
	
	\caption{The 3-D surface morphology of CsI thin films obtained by AFM technique in case of "as-deposited" (a, c, e, g) and "1 hour humid air aged" (b, d, f, h).}
\end{figure}

Surface morphology plays an important role in various fields of modern day science and technology. It exhibits rough behavior on a nanometer scale; hence its characterization is very necessary.  Generally, interface width ($\sigma$) is the measure of height fluctuation of the surface, which is defined as $\sigma= \frac{1}{N}\sqrt{{\langle h(i,j) - \overline{h(i,j)} \rangle}}^{2}$ , where $h(i,j)$ is the height distribution of surface at points $(i,j)$  and $\overline{h(i,j)}$ is overall average height. The calculated values of $\sigma$ for "as-deposited" and "1 hour humid air aged" films are 0.0404, 0.0156, 0.0156, 0.0233, 0.0194 $\mu m$  and 0.0371, 0.0436, 0.0245, 0.0613 $\mu m$ respectively. The value of $\sigma$ is only sensitive to the peak and valley of surface profile but does not give any information about correlation and irregularity/complexity of a surface.  For this purpose fractal analysis is highly needed. 

At the first sight, thin film morphology appears to be random; a close analysis reveals the values of correlation properties. The best way to probe correlation property of thin film surface is described by autocorrelation function, $A(r)$ as well as by height-height correlation function, $H(r)$ . The normalized autocorrelation function is described by, $A(r= md)= \frac{\sigma^{-2}}{N(N-m)} \langle [h(i+m,j) h(i,j)] \rangle$~\cite{2016_120_5755−5763}, here $N$ is the total number of pixels and $m$ refers to the current data points or pixels used in calculations while $d$ is the horizontal distance between two adjacent pixels. We plotted a graph between $A(r)$ and $r$ as for "as-deposited" and "1 hour humid air aged" films which is depicted in Figs. 6a and 6b. The value of $r$ at which $A(r)$ drops to $1/e$ of its original value is defined as the lateral correlation length $\xi$. In other words, $A(r)$ can be used to detect non-randomness in surface data as well as to identify an appropriate space if the data are not random. It provides vertical as well as lateral information about the surfaces~\cite{347_(2015)_706–712}. From figure, we can see that "as-deposited" thin films surfaces have quasi-periodic nature while after 1 hour humid air aged the amplitude of quasi-periodic nature (vertical portion) reduced (become lateral).  For more details, compare lateral correlation length from Tables 3 and 4. It is also observed that maximum kinks/spikes are removed after exposing to humidity. The calculated values of $\xi$ for "as-deposited" and "1 hour humid air aged" thin films are listed in Tables 3 and 4.  From table, it is clear that "1 hour humid air aged" films have larger $\xi$ than that in "as-deposited" films.

The height-height correlation function has the scaling form and mathematically given by\\ $H(r)= \frac{1}{N(N-m)}\langle [h(i+m,j) - h(i,j)]^2 \rangle$~\cite{347_(2015)_706–712}. The height-height correlation function is also very intimately associated with the autocorrelation function and is described by the relation $H(r)= 2\sigma^{2}[1-A(r)]$~\cite{122_185303_(2017)}. A graph is plotted between $log~H(r)$ vs $log~r$ for each "as-deposited" and "1 hour humid air aged" thin film as shown in Figs. 6c and 6d. From the figure, we see that two distinct regions are clearly observed and each surface shows oscillatory behavior for sufficiently large value of $r \gg \xi$. It is expected that for small distance $r$, $log~H(r)$ should increase linearly with $log~r$, depicting a power-law behavior represented by $H(r) \sim r^{2\alpha}$, where $\alpha$ is known as roughness exponent ($\alpha$). It is obtained from a fit to the linear portion of $log~H(r)$ versus $log~r$ plot. At the initial regime of the correlation function there exist variant slopes representing different micro-structural evolutions in "as-deposited" and "1 hour humid air aged" thin films. The value of $\alpha$ signifies the how "wiggly" the local slope is. It is also an indicative of the "jaggedness" of surface morphology and lies between 0 and 1. The smaller value of $\alpha$ correspond to more wiggly surface. The computed value of $\alpha$ for each thin film is listed in Tables 3 and 4.

On the very short length scale, the parameter $\alpha$ is connected with the fractal dimension as $D_f= 3-\alpha$~\cite{347_(2015)_706–712}. The value of $D_f$ is the measure of two- and three-dimensional thin film growth mechanism.  Therefore, fractal dimension is one of the most informative parameter of surface geometry. The higher value of fractal dimension is corresponding to more irregular/complex surface. The computed value of $D_f$ for each thin film is also listed in Tables 3 and 4.

\begin{figure}
	\centering
	\includegraphics[width=\columnwidth,height=5cm]{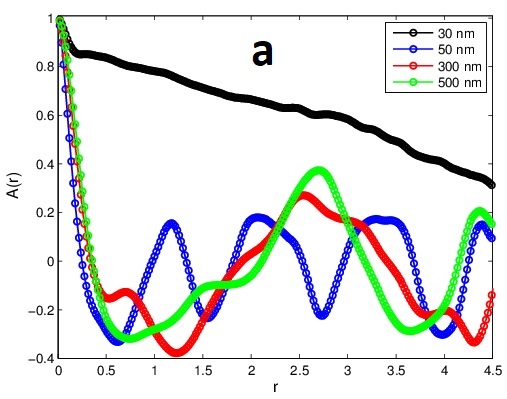}
	\vspace{0.5cm}
	\includegraphics[width=\columnwidth,height=5cm]{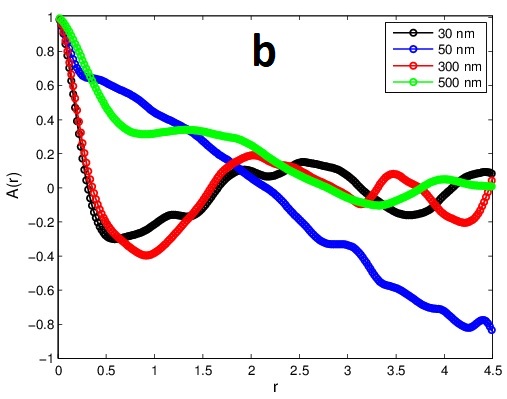}
	\vspace{0.5cm}
	\includegraphics[width=\columnwidth,height=5cm]{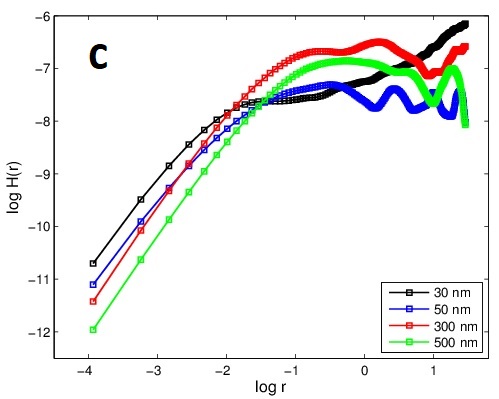}
	\vspace{0.5cm}
	\includegraphics[width=\columnwidth,height=5cm]{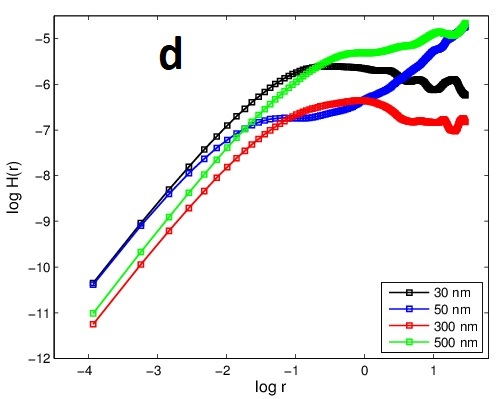}
	\caption{The Autocorrelation function A(r) versus r and the log H as a function of log r for different thicknesses of CsI thin films in case of: (a, c) "as-deposited" and (b, d) "1 hour humid air aged".}
\end{figure}	

\begin{table}
	\begin{center}
		\caption{The fractal parameters of "as-deposited" CsI thin films deposited on Stainless Steel substrate.}	 
		~\\   
		\begin{tabular}{c c c c c} 
			\hline
			t & $\sigma$ & $\xi$ & $\alpha$ & $D_f$ \\
			[0.5ex]
			(nm) & (nm) & (nm) & ~ & ~
			\\[0.5ex]
			\hline
			30	& 0.0404 & ... &   0.7649 & 2.2351 \\
			50	& 0.0156 & 0.1467 & 0.7135 & 2.2865 \\		
			300	& 0.0233 & 0.2079 & 0.8744 & 2.1256\\		
			500	& 0.0194 & 0.2314 & 0.8925 & 2.1075\\		
			\hline
			
		\end{tabular}
	\end{center}
\end{table}

\begin{table}
	\begin{center}
		\caption{The fractal parameters of "1 hour humid air aged" CsI thin films deposited on Stainless Steel substrate.}	 
		~\\   
		\begin{tabular}{c c c c c} 
			\hline
			t & $\sigma$ & $\xi$ & $\alpha$ & $D_f$ \\
			[0.5ex]
			(nm) & (nm) & (nm) & ~ & ~
			\\[0.5ex]
			\hline
			30	& 0.0371 & 0.2025 & 0.8577 & 2.1423 \\
			50	& 0.0436 & 1.2575 & 0.8390 & 2.1610 \\		
			300	& 0.0245 & 0.2222 & 0.8529 & 2.1471\\		
			500	& 0.0613 & 0.6490 & 0.9109 & 2.0891\\		
			\hline
			
		\end{tabular}
	\end{center}
\end{table}

\section{\label{sec:level1}Conclusion}
The variation of grain size obtained from TEM technique has been found to range from $\sim$ 313 nm to $\sim$ 1058 nm. The average grain size has been found to increase after exposing to humidity. It is increased from $\sim$ 679 nm to $\sim$ 724 nm. The increase in average grain size due to the humidity is attributed to the coalescence process. The micro-structure results obtained from SAED pattern show that the appeared lattice planes are attributed to bcc structure with lattice constant about 4.56 $\angstrom$. The experimental values of inter-planner spacing are found to be less than the standard values which implies a compressive stress acting in the films. Further, the elemental compositions of 100 nm thick CsI film has been investigated by the means of EDAX technique. It is found that the atomic percentage ratio of Cs to I is $\sim$ 1:1 in both cases. The atomic percentage has been increased by two times after exposing to humidity.

The AFM images of the "as-deposited" and "1 hour humid air aged" CsI thin films surface morphologies have been analyzed using fractal techniques. The autocorrelation function and the height-height correlation functions are used for this purpose. We quantify the lateral correlation length, roughness exponent, interface width, and the fractal dimension for both cases. The value of lateral correlation length is increased in each case after exposing to humidity. The smaller values of roughness exponent correspond to wigglier surface while larger one indicates the smoother surface.  The fractal dimension provides information about the variations in the surface morphology. A larger value of fractal dimension corresponds to more jagged morphology, while a smaller one indicates locally smooth surface structure.

\section*{Acknowledgment}

This work was partially supported by the FIST and PURSE programs of Department of Science and Technology (DST), CAS and UPE programs of University Grant Commission (UGC) and by the Indian Space Research Organization (ISRO) SSPS program, Government of India.

\end{document}